\documentclass[twocolumn,twoside,slac]{revtex4}
\usepackage{graphicx}
\usepackage{fancyhdr}
\newcommand{\be}{\begin{equation}}
\newcommand{\ba}{\begin{eqnarray}}
\newcommand{\ee}{\end{equation}}
\newcommand{\ea}{\end{eqnarray}}  

\def\gtsima{$\; \buildrel > \over \sim \;$}
\def\ltsima{$\; \buildrel < \over \sim \;$}
\def\gsim{\lower.5ex\hbox{\gtsima}}
\def\lsim{\lower.5ex\hbox{\ltsima}}
\def\simgt{\lower.5ex\hbox{\gtsima}}
\def\simlt{\lower.5ex\hbox{\ltsima}}
\def\simpr{\lower.5ex\hbox{\prosima}}

\def\msun{{M_\odot}}

\def\E3{{\cal E}_{\rm g}^{III}}

\begin{document}

\title{Temporal Bias in the Clustering of Massive Cosmological Objects}

%

\author{Evan Scannapieco}
\affiliation{Kavli Institute for Theoretical Phys., Kohn Hall, UC
Santa Barbara, Santa Barbara, CA 93106}
\author{Robert J. Thacker}
\affiliation{Dept.\ of   Phys.\ \& Astron.,
McMaster Univ., 1280 Main St.\ West, Hamilton, Ontario, L8S 4M1,
Canada}

\begin{abstract}

It is a well-established fact that massive cosmological objects
exhibit a ``geometrical bias'' that boosts their spatial correlations
with respect to the underlying mass distribution.  Although this
geometrical bias is a simple function of mass, this is only half of
the story. We show using numerical simulations that objects that are
in the midst of accreting material also exhibit a ``temporal bias,''
which further boosts their clustering far above geometrical bias
levels. These results may help to resolve a discrepancy between
spectroscopic and clustering mass estimates of Lyman Break Galaxies, a
population of high-redshift galaxies that are caught in the act of
forming large numbers of new stars.

\end{abstract}

\maketitle

\thispagestyle{fancy}


\section{Introduction}

Large-scale structure in the Universe is believed to have originated
from a primordial Gaussian random field of matter fluctuations, the
product of quantum fluctuations that were shifted to larger spatial
scales during cosmological inflation [{\em e.g.\ } 1].  Cosmic
Microwave Background observations show that when the Universe was
100,000 years old, the gaseous component was extremely smooth, with
temperature variations $\sim 10^{-5}$.  These tiny inhomogeneities, in
concert with inhomogeneities in the unseen massive dark-matter, were
amplified through gravitational instability, eventually forming the
galaxies, clusters, and other cosmological objects we see today.

While the precise details of structure formation are highly complex, the
simplicity of the initial random field allows us to easily compute the
overall distributions of cosmological objects.  Galaxies and galaxy
clusters represent the peaks in the initial density distribution which
accrete matter at the expense of the diffuse regions between them, and
thus their number densities can be simply related to the number densities
of the peaks of the initial random field.  This technique has been applied
most cleanly to galaxy clusters, whose densities and evolution provide
strong constraints on the overall matter density \cite{ek96}.

Similarly, because the peaks in a random field are more clustered than
the overall distribution, the spatial clustering of cosmological
objects is stronger than the underlying mass distribution.
Furthermore, this ``geometrical bias'' is a systematic function of the
mass of these structures, an effect that has been well-studied
analytically and numerically [{\em e.g.\ } 3, 4, 5].

Yet, this is only half of the story.  Here we conduct a detailed
numerical simulation that shows that the spatial correlation function
of objects that are in the midst of accreting substantial amounts of
material is significantly enhanced over that of the general
population.  This temporal bias causes them to mimic the properties of
higher-mass structures, with important astrophysical implications as
discussed below.

The structure of this work is as follows: In \S 2 we describe our
numerical simulation, discuss our group-finding algorithms, and
develop a robust definition of accreting groups.  In \S 3 we present
our results for the spatial correlation functions of these samples,
and in \S 4 we discuss the astrophysical implications of our results.
Further details of this study are given in \cite{sc03}.

\section{Simulations and Group Finding}

Our numerical simulation traced the growth of primordial density
fluctuations by dynamically evolving a large number of point test
particles.  The distribution of these particles was then used to
determine the nonlinear evolution of the spatial correlation function
of massive objects at late times.  Driven by measurements of the
Cosmic Microwave Background, the number abundance of galaxy clusters,
and high redshift supernova estimates [{\em e.g.\ }\ 7, 2, 8] we
focused our attention on a Cold Dark Matter cosmological model with
parameters $H=70$ km/s/Mpc, $\Omega_0$ = 0.3, $\Omega_\Lambda$ = 0.65,
$\Omega_b = 0.05$, and $\sigma_8 = 0.87$, where $H$ is the Hubble
constant today, $\Omega_0$, $\Omega_\Lambda$, and $\Omega_b$ are the
total matter, vacuum, and baryonic densities in units of the critical
density, and $\sigma_8^2$ is the present variance of linear
fluctuations on the $8\times(100/70)$ Mpc scale (where 1 Mpc is $3.26
\times 10^6$ light years).

Periodic boundary conditions, which approximate large-scale
homogeneity and isotropy, were taken, and the mass within our
simulation volume was held fixed.  Thus the overall box expanded along
with the cosmological expansion, such that each side at any given
redshift $z$ was $73/(1+z)$ Mpc across.  This box was populated with
$350^3$ dark matter particles that interact only gravitationally and
represent the dominant mass component of the Universe.  The mass of
each particle, $4.3 \times 10^{8}$ solar masses ($\msun$), was chosen
to match the observed mass density of the Universe, and the simulation
was started at an initial redshift of $z=49$.  The simulation used a
parallel OpenMP-based version of the HYDRA code [9, 10] with 64-bit
precision.


To demonstrate the robustness of our results we have chosen two
distinct group-finding approaches; the friends-of-friends approach
\cite{da85} (FOF) and the HOP algorithm \cite{ei98}.  FOF works by
linking together all pairs of particles within a fixed
``linking-length'' of each other, and then taking each such group of
``friends'' to be an identified cosmological object.  Although it
remains popular, the FOF mass estimates are known to have
significant scatter due to a problem that can occur as small strings
of particles fall within the linking length.

The HOP algorithm works by using the local density for each particle
to trace (`hop') along a path of increasing density to the nearest
density maxima, at which point the particle is assigned to the group
defined by that local density maximum. As this process assigns all
particles to groups, a `regrouping' stage is needed in which a merger
criterion for groups above a threshold density $\delta_{outer}$ is
applied. This criterion merges all groups for which the boundary
density between them exceeds $\delta_{saddle}$, and all groups thus
identified must have one particle that exceeds $\delta_{peak}$ to be
accepted as a group (see \cite{ei98} for explicit details).

Beginning from $z=4.89$, we saved particle positions every 50 million
years up to the final output at $z=3$. For the final 5 outputs we
found FOF groups using a linking parameter of $b=0.18$, and HOP groups
using the parameters: $N_{dens}=48$, $N_{hop}=20$, $N_{merge}=5$,
$\delta_{peak}=160$, $\delta_{saddle}=140$, and
$\delta_{outer}=80$. Visual inspection showed strong similarities
between the two populations, with a small amount of unavoidable noise
coming from groups around the 80 particle resolution limit (a group
found by FOF at this limit may not be found by HOP and vice versa).
We compare groups from one output to another by tracing back all
particles with a given group index from the later output to the
earlier output.  Particles that show no membership to a group at the
earlier time are regarded as `smooth infall' while other non-null
indices describe the merger history of the object.

To give a rough estimate of the accuracy of the group finding methods
in Fig.\ 1 we plot the mass of the most massive progenitor at
$t_1(z=3.059)$, versus the mass at $t_2(z=3)$, such that $\Delta t = 5
\times 10^7$ years.  As compared to the FOF groups, the smaller
fraction of HOP groups lying above the equal mass line shows that the
HOP algorithm identifies groups that are more likely to be massive at
later outputs.  The effect of this difference is significant.

Our definition of accreting groups is similar to that of \cite{pe03},
except that we select the subset that grew by 20\% from $t_1$ to
$t_2$, which implicitly includes mass accretion via smooth infall and
results in 545(980) HOP(FOF) groups if $\Delta t = 5 \times 10^7$
years. Note that the mass of each group is that at the end of each
time interval, such that we tag all groups that {\em experienced}
appreciable infall.  The 20\% value is arbitrary, but we selected it
primarily because it appears to lie outside the central `noise' band
in the FOF data (see Fig\ 1). The groups corresponding to this cut
appear in Fig\ 1, as points to the right of the dashed lines.

\begin{figure}
\includegraphics[width=65mm]{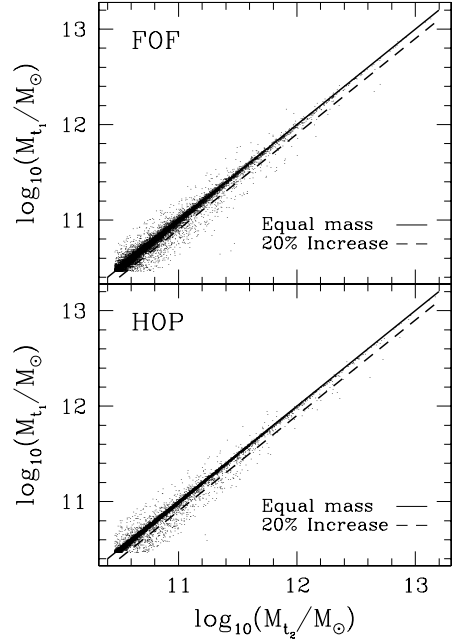}
\caption{Comparison of group
growth. The FOF algorithm exhibits a significant amount of scatter in mass
estimates between outputs. Only 67\% of groups grow from time $t_1$ to 
$t_2$, compared with 82\% for HOP.} 
\end{figure}

\section{Temporal Bias}

\setcounter{figure}{1}
\begin{figure*}[t]
\vspace{114mm}
\includegraphics{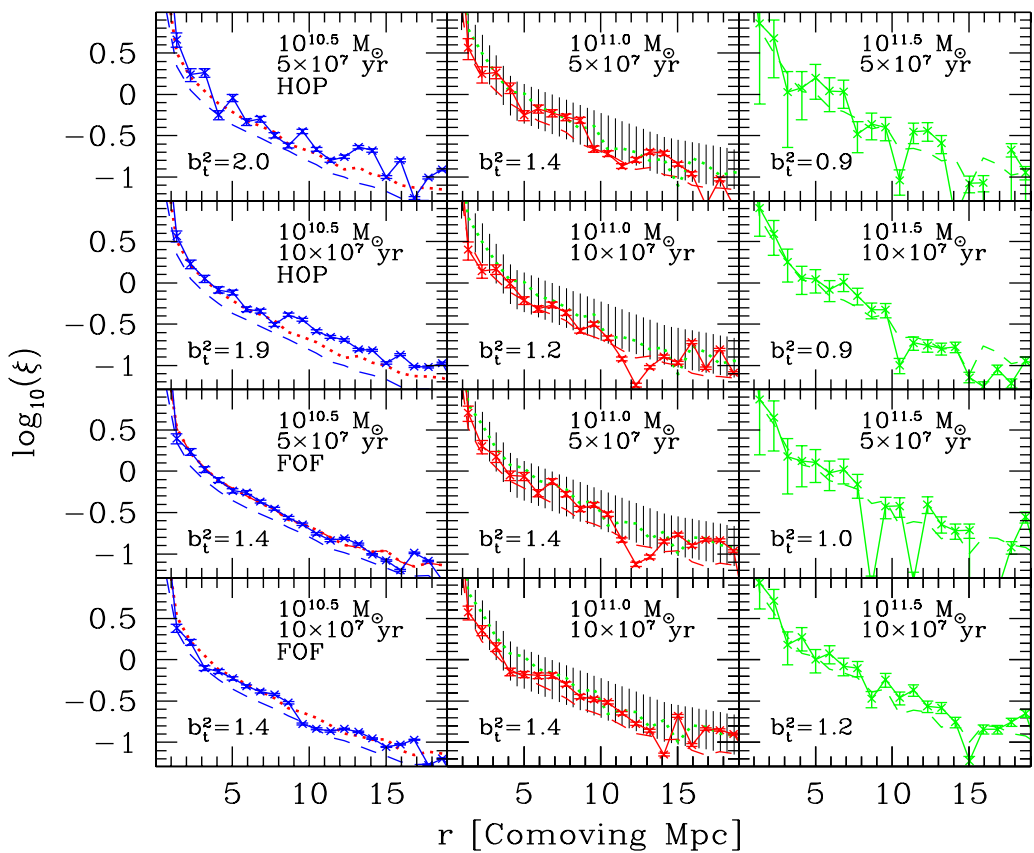}
\caption{Spatial Correlation Functions.
In each panel the dashed line shows the correlation function for all
the groups, while the points connected by the solid line show
$\xi(r)$ for groups that have accreted appreciable mass in the last
$\Delta t$ years.  Panels are labeled by their mass range and 
$\Delta t$ values, and in each panel, the dotted line shows the
correlation function of all the groups in the next highest mass bin.
The top two rows were generated from a set of groups selected by the
HOP algorithm, while the groups in the lower two rows were selected
using the FOF approach.  The shaded region in the central panels
represents the observed correlation function of ${\cal R}_{AB} \leq 25.5$
Lyman break galaxies as computed in \protect\cite{we01} by inversion
of the angular correlation function.  A 10\% accretion threshold is applied
in the $10^{11.5} \msun$ case to increase the number of measured groups.}
\label{fig:xi}
\end{figure*}

In Figure \ref{fig:xi} we show the spatial correlation functions of the
groups selected by both the HOP and FOF algorithms and compare them
with $\xi(r)$ of the accreting groups.  This function measures the
excess probability of finding a pair of groups at a given separation
$r$ relative to a random distribution, and is calculated for a separation 
bin $r_l$ as 
\be
1 + \xi(r_l) = N(r_l)/N_{\rm random}(r_l),
\ee
where $N(r_l)$ is the number of pairs separated by distances between
$r_l$ and $r_{l+1}$, and $N_{\rm random}(r_l) = \frac{1}{2} N^2 \frac{4 \pi}{3}
({r_{l+1}^3 - r_l^3})/V$, with $N$ the total number of groups and $V$
the volume of the simulation.
In the accreting case we co-added the correlation functions calculated
from the differences from the last four $\Delta t = 5 \times 10^7$
year intervals and the last two $\Delta t = 10 \times 10^7$ year
intervals. Radial bins of 1/80 the simulation size, corresponding to
$0.92$ comoving Mpc, were taken throughout.  For comparison, in each
panel of Fig.\ \ref{fig:xi} we also show the correlation function of
all the groups in the next largest mass bin.  The amplitudes of the
correlation functions obtained using the full set of HOP and FOF
groups agree with each other to within statistical uncertainties, as
well as with analytical estimates.

The upper row demonstrates a clear enhancement of the clustering of
accreting groups at both the $10^{10.5} M_\odot$ and $10^{11.0}
M_\odot$ mass scales, with their correlation functions roughly
matching those of objects three times greater in mass (no conclusion
can be drawn from the high mass bin as the sample is too small).  This
``temporal biasing'' arises from the fact that {\em both} objects
accreting substructure as well as those experiencing considerable
smooth infall tend to be found in the densest regions of space, which
are themselves highly clustered. This conclusion is supported by the
fact that the average local overdensity of groups in the
$10^{10.5}(10^{11}) M_\odot$ mass bin is 0.82(0.87) (measured in 4 Mpc
(comoving) spheres, corresponding to a mass scale of
$1.2\times10^{13}$ $M_\odot$), whereas the same mass bin for the
entire population exhibits an overdensity of 0.60(0.73).

In the second row of Fig.\ \ref{fig:xi} we take a longer interval of
$\Delta t = 10 \times 10^7$ yr.  This has a only slight dampening
effect on temporal bias, which can not be definitively distinguished
from statistical noise in our measurements.  In the $\Delta t = 20
\times 10^7$ yr case, however, only a very weak enhancement of
$\xi(r)$ was measured.

In the lower two rows of this figure, we repeat our analyses using the
FOF group finder.  Although this approach is more susceptible to
statistical noise, the same trends are apparent as in the HOP case.
If $\Delta t = 5 \times 10^7$ yr, this temporal bias is roughly equal
to the geometrical bias of the groups three times more massive, while
if $\Delta t = 10 \times 10^7$ yr, $\xi(r)$ is boosted to a slightly
lesser degree.

Finally, to quantify our results, we have computed the effective
temporal bias in each mass bin, $\Delta t$, and group finder.  We
define $b^2_t$ as the ratio of the correlation function of the
accreting groups to the overall correlation function, weighted by the
number of points in each bin in the overall function; $b^2_t \equiv
\sum^{20}_{i=0} \frac {\xi_{{\rm accreting},i} N_{{\rm
all},i}}{\xi_{{\rm all},i}N_{{\rm all},i}}$, where the sum is carried
out over all bins within $r \leq 20$ comoving Mpc.  These values are
labeled in each panel, and in the $\Delta t = 20 \times 10^7$ yr case,
$b^2_t =$ 1.1(1.0) in the $10^{10.5}(10^{11.0}) \msun$ HOP bins and 
1.1(1.3) in the respective FOF bins.

\section{Astrophysical Implications}

While temporal bias is a general property of the peaks of a
gravitationally amplified Gaussian random field, our results have
specific implications for the large sample of $z \sim 3$ galaxies made
available by the Lyman-break color-selection technique \cite{st98}.
Lyman break galaxies (LBGs) are observed to have enormous
star-formation rates on the order of $\sim 50 \msun$ per year
\cite{ad00}, implying that these objects are likely to be accreting
large amounts of material.  Furthermore, although the clustering of
LBGs brighter than ${\cal R}_{AB} \leq 25.5$ is roughly that expected
from the geometrical bias of $10^{12} \msun$ objects [{\em e.g.\ }16,
17], the linewidths measured from a spectroscopic sample these
galaxies correspond to total masses $\leq 10^{11} M_\odot$
\cite{pe01}.

To relate our result to LBGs we plot the spatial correlation function
of ${\cal R}_{AB} \leq 25.5$ LBGs, as derived in \cite{we01}, in the
center column of Fig.\ \ref{fig:xi}. Although there are significant
uncertainties involved in computing this quantity, since comparisons
are more naturally conducted in angular coordinates, the shaded
regions provide a guide to the range of $\xi(r)$ values consistent
with observations.  In these panels, we see that if $\Delta t = 5
\times 10^7$ yr is chosen, then temporal bias boosts the correlation
function of $10^{11} \msun$ groups into reasonable agreement with
observations.

This mass is marginally consistent with the upper mass bound inferred
from the rotation curves of a somewhat bright (${\cal R}_{AB} \lesssim
24$) spectroscopic subset of LBGs \cite{pe01}.
Furthermore, only $\sim 4\%$ of all groups exhibit appreciable
accretion in each $\Delta t = 5 \times 10^7$ year time interval and
the density of $10^{11} \msun$ groups is $\sim 2 \times 10^{-2}$
Mpc$^3$, at $z = 3$ in our assumed cosmology.  Thus associating such
objects with $5 \times 10^7$ year starbursts results in a density
$\sim 5 \times 10^{-4}$ Mpc$^3$, comparable with that observed.

While quite suggestive, these comparisons are not meant as a complete
model, and may not prove to be the final explanation of the discrepant
mass estimates of LBGs.  Kinematic models have been explored, for
example, in which the observed velocity dispersions of LBGs are much
less than the circular velocities of the groups in which they are
contained \cite{mo99}.  What is clear however, is
that this bias can not be ignored and must be carefully considered
when interpreting the clustering of these objects.  While perhaps only
part of the story, temporal biasing represents an important factor
that must be taken into account when studying the properties of Lyman
break galaxies.

\begin{acknowledgments}

ES would like to express his sincere thanks for the hospitality shown
to him by Jon Weisheit and the T-6 group at Los Alamos National
Laboratory, where this work was initiated.  We are grateful to Marc
Davis for fruitful suggestions.  ES was supported in part by an NSF
MPS-DRF fellowship.  RJT\ acknowledges funding from the Canadian
Computational Cosmology Consortium and use of the CITA computing
facilities.  This work was supported by the National Science
Foundation under grant PHY99-07949.

\end{acknowledgments}


\begin{thebibliography}{}

\bibitem{li83}
Linde, A. D. 1983, Phys. Lett., 129B, 177
\bibitem{ek96}
Eke, V. R., Cole, S., \& Frenk C. S. 1996, Monthly Notices of the Royal Astronomical Society, 282, 263
\bibitem{ka84} 
Kaiser, N. 1984, Astrophysical Journal, 284, L9
\bibitem{mo96}
Mo, H. J. \& White S. D. M., 1996, Monthly Notices of the Royal Astronomical Society, 282, 348
\bibitem{ji99}
Jing, Y. P. 1999, Astrophysical Journal, 515, L45
\bibitem{sc03} 
Scannapieco, E. \& Thacker, R. J. 2003, Astrophysical
Journal Letters 590, 69
\bibitem{sp03} 
Spergel, D. N. et al.\ 2003, Astrophysical Journal Supplement,
148, 175
\bibitem{pe99} 
Perlmutter, S. et al.\ 1999, Astrophysical Journal, 517, 565
\bibitem{co95} 
Couchman, H. M. P., Thomas, P. A., \& Pearce, F. R. 1995, Astrophysical Journal, 452, 797
\bibitem{th00} 
Thacker, R. J \& Couchman, H .M. P. 2000, Astrophysical Journal, 545, 728
\bibitem{da85} 
Davis, M., Efstathiou, G., Frenk, C. S., \& White,
S. D. M. 1985, Astrophysical Journal, 292, 371
\bibitem{ei98}
Eisenstein, D. J. \& Hut, P. 1998, Astrophysical Journal, 498, 137
\bibitem{pe03}
Percival, W. J., Scott, D., Peacock, J., A., \& Dunlop, J. S.
	2003, Monthly Notices of the Royal Astronomical Society, 338, L31 
\bibitem{st98} 
Steidel, C. C., Adelberger, K. L., Dickinson, M., Giavalisco,
  M., Pettini, M., Kellogg, M. 1998, Astrophysical Journal, 492, 428
\bibitem{ad00} 
Adelberger, K. L. \& Steidel, C. C. 2000, Astrophysical Journal, 544, 218
\bibitem{we01} 
Wechsler, R. H., Somerville, R. S., Bullock, J. S.;
  Kolatt, T. S.; Primack, J. R.; Blumenthal, G. R.; Dekel,
  A. 2001, Astrophysical Journal, 554, 85
\bibitem{po02}
Porciani, C. \& Giavalisco, M. 2002, Astrophysical Journal, 565, 24
\bibitem{pe01}
Pettini, M et al.\ 2001, Astrophysical Journal, 554, 981
\bibitem{mo99}
Mo, H. J., Mao, S., \& White, S. D. M. 1999, Monthly Notices of the Royal Astronomical Society, 304, 175

\end{thebibliography}
\end{document}